# Initiation and spread of escape waves within animal groups

James E. Herbert-Read[1,2,†], Jerome Buhl[3,4,5,†], Feng Hu[6], Ashley J. W. Ward[3] and David J. T. Sumpter[1,2]

[1]Department of Mathematics, and [2]Department of Ecology and Genetics, Uppsala University, Uppsala 75106, Sweden
[3]School of Biological Sciences, and [4]The Charles Perkins Centre, The University of Sydney, Sydney, New South Wales 2006, Australia
[5]School of Agriculture, The University of Adelaide, South Australia 5005, Australia
[6]College of Physics and Electronic Engineering, Chongqing Normal University, Chongqing City 400047, China

## 1. Summary

The exceptional reactivity of animal collectives to predatory attacks is thought to be owing to rapid, but local, transfer of information between group members. These groups turn together in unison and produce escape waves. However, it is not clear how escape waves are created from local interactions, nor is it understood how these patterns are shaped by natural selection. By startling schools of fish with a simulated attack in an experimental arena, we demonstrate that changes in the direction and speed by a small percentage of individuals that detect the danger initiate an escape wave. This escape wave consists of a densely packed band of individuals that causes other school members to change direction. In the majority of cases, this wave passes through the entire group. We use a simulation model to demonstrate that this mechanism can, through local interactions alone, produce arbitrarily large escape waves. In the model, when we set the group density to that seen in real fish schools, we find that the risk to the members at the edge of the group is roughly equal to the risk of those within the group. Our experiments and modelling results provide a plausible explanation for how escape waves propagate in nature without centralized control.

## 2. Introduction

Many animal collectives such as fish schools and insect swarms are highly responsive to perturbations such as predatory attacks [1–3]. It has been proposed that this is owing to the group self-organizing into particular spatial configurations [4–7]. Information about a threat can then be transmitted through the body movements of individuals responding both directly to the threat and indirectly to the movements of neighbours [8].

One resulting pattern is a wave of individuals turning away from the threat. The distance this wave travels can be thought of in terms of information transfer, with information about the threat propagating through members of the group. Studies of fish schools and bird flocks have shown that the distance information travels scales with group size [9,10]. In starling flocks and schooling fish, for example, the information travels without substantial dampening through hundreds or thousands of individuals [10,11]. In some cases, the way that individuals in a group turn away from a threat [12] is much like how information is propagated through bird flocks during collective turns [11].

Escape waves are less well understood at the individual level. Previous work has demonstrated that startled fish in groups have longer response latencies than individuals on their own [13]. Furthermore, schooling fish show more directed movements away from threats compared with solitary individuals [13,14], and have shorter response latencies when closer to a simulated threat than when further away from it [13]. Because individuals closest to a threat are quicker to initiate escape responses compared with those further away [15], these individuals may provide information that can trigger an escape wave throughout the group. While these experiments provide useful information about the response of fishes to threats, the mechanism used to initiate and propagate escape waves remains unclear.

One important way in which animals respond to each other's movements, and which could allow effective information transfer, is through changes in speed [9,16–18]. Individuals use the speed changes of neighbours to inform their own movements [19,20]. Because individuals in groups increase their speed in response to predatory attacks [1,21,22], speed cues could indirectly inform neighbours about the location of a threat. Indeed, small fluctuations in the velocities of individuals can be correlated not only locally, but also at large distances between individuals, implying information about a threat can spread across entire groups [23]. We investigate how these speed changes may be used to initiate escape waves.

Another question is whether a group's spatial configuration facilitates information transfer. Simulation models suggest that there is a range of densities where directional information propagates more efficiently [8]. At very low group densities, there are insufficient interactions to produce coordinated motion, whereas at very high densities, motion becomes highly coordinated, but does not respond to perturbations. This question becomes more difficult to answer when we consider the different risks associated with positioning within the group [24,25]. Individuals who position themselves in the centre of the group are generally further away from predators and may be less vulnerable than those individuals at the margins of the group [26,27]. However, the rapid reorganization of spatial positions following a predatory attack may imply that the periphery is not more dangerous than the centre [28–30].

To investigate these ideas, we studied the evasion response of fish schools (Pacific blue-eyes, *Pseudomugil signifer*) to a simulated predatory attack. Pacific blue-eyes are a facultative schooling species [31] hunted by both fish and avian predators. Groups of fish were placed into an annulus arena where they formed a polarized school travelling around the arena. We perturbed the schools using a paddle-like stimulus that was extended above the surface of the water towards the group's leading edge, causing fish to turn 180° in an attempt to evade the threat. Using a combination of automated tracking software and self-propelled particle modelling, we investigated whether the changes in speeds of a minority of group members following the attack could allow the group as a whole to avoid the threat. We also determined the relative costs and benefits for individuals occupying different spatial positions within the group.

## 3. Material and methods

### 3.1. Study species and experimental methods

Approximately 340 Pacific blue-eyes (*P. signifer*) were caught in hand nets from Narrabeen Lagoon, New South Wales, Australia (33°43′03 S, 151°16′17 E). All collection abided by the NSW Department of Primary Industries §37 Permit. Fish were kept in filtered freshwater in 150 l glass tanks and fed crushed flake food ad libitum. Fish were approximately 2–3 cm standard length. All fish were housed for at least three weeks prior to experimentation.

We constructed an experimental annulus arena (760 mm external diameter, 200 mm internal diameter) and filled it to a depth of 70 mm with aged and conditioned tap water. A camera (Logitech Pro 9000) placed directly above the centre of the arena filmed the experiments at 15 frames per second. The arena was lit by fluorescent lamps and was visually isolated.

For each experimental trial, we randomly selected individuals from the housing tanks and placed them in the experimental arena. We selected a different number of fish for each trial, ensuring that we had representative samples of 10–158 individuals for our replicates ($n = 39$). When placed into the arena, fish formed a polarized group swimming around the annulus arena. Fish were left to acclimate to the arena for at least 2 min, after which we waited until the fish were swimming in a clockwise (CW) direction. The stimulus used to perturb the fish schools was a $6\,cm^2$ piece of opaque black plastic fixed to the end of a white rod, 4 mm in diameter, that could be horizontally extended 200 mm out and above the surface of the water (at a height of 2–3 cm above the water's surface). When a fish was within 150 mm of the stimulus, we extended the stimulus above the water in the opposite direction of group travel. This generally caused the fish to turn around and swim in an anticlockwise (anti-CW) direction (electronic supplementary material, movie S1). The mean time for the stimulus to fully extend into the arena was $0.42\,s \pm 0.18\,s$ ($\pm 2$ s.d.). Note that this is the time from when the stimulus started to move to when it had stopped moving and was fully extended above the surface of the water. Any trials that did not fall within this range were discarded, because the escape behaviour of fish can be affected by the speed of an approaching predator [32]. The same protocol described above was used for experiments with single fish ($n = 30$). In these experiments, however, only single fish were selected for trials, and not groups. Apart from pack hunting, most fish predatory–prey interactions involve a stalking event followed by a lunging attack [33], which is why we selected this form of stimulus. We standardized the direction of the attack but note that it could vary in natural predatory–prey interactions. Our experimental set-up should not have substantially constrained the escape movements of individual fish, because in other more open experimental set-ups, the escape trajectories of fish are still often directed away (140–160°) from the direction of attack [13]. We do note that varying the direction of attack on schools can effect the immediate alignment of their evasive manoeuvres (within 100 ms of the attack); however, schools regain alignment and travel in the opposite direction of attack after 500 ms [34]. The tank provided sufficient space for a shoal to form with a density similar to that observed in nature. In particular, the fish typically aggregated to occupy only a portion of the available space.

## 3.2. Data collection

Films were converted from .wmv format to .avi using DIRECTSHOWSOURCE and VIRTUALDUB (v. 1.9.2) and were subsequently tracked using DIDSON by Handegard & Williams [35], giving the raw $x$-, $y$-coordinates of each fish at each frame. Two detected objects, $i$ and $j$, on frames $t$ and $t-1$ respectively, were considered to be the same fish if $j$ was the nearest neighbour to $i$ and reciprocally; otherwise, the object $i$ was discarded. If $i$ was tracked properly, then its orientation and speed were estimated using the angle and module of the vector $ji$. To check the accuracy of the tracking process, the resulting tracking data were plotted as an overlay to experiment frames, and the resulting videos for all experiments were checked for tracking errors and artefacts. On average, we tracked 87% ($\pm 13$ s.d.) of fish across all frames of all experiments. The percentage of frames in which over 80% of fish were tracked was 79%. This is a similar level of accuracy as other recent studies [4].

## 3.3. Data analysis

For each fish, we defined its relative orientation as $\chi = \arcsin(\sin(\theta - \vartheta))$, where $\theta$ was the angle of the fish and $\vartheta$ is the angle of the arena radius going through the fish position. A fish moving parallel to the radius of the arena (either in a centripetal or centrifugal way) will thus have a relative orientation $\chi = 0$, whereas a fish perpendicular to the radius will have a relative orientation of $\chi = -\pi/2$ if moving CW and $\chi = \pi/2$ if moving anti-CW. We then divided the arena into 36 equal sectors. In each of these sectors at each timestep $t$, we calculated the instantaneous alignment $\phi_{x,t}$ as the average of the orientation for all moving fish as follows:

$$\phi_{x,t} = \frac{2}{|m_{x,t}|\pi} \sum_{i \in m_{x,t}} \chi_{i,t},$$

where $m_{x,t} = \{i : x < x_{i,t} < x + (2\pi/36)\}$ is the number of fish within the sector at distance $x$ from the stimulus; $x_{i,t}$ is the position of fish $i$ at time $t$, and $\chi_{i,t}$ its angle. Thus, values of the alignment close to the extreme values of 1 and $-1$ indicate all fish facing towards, or respectively away from, the stimulus. Values close to zero indicate an absence of any collective alignment (electronic supplementary material, figure S1). Average local density was calculated by dividing the number of radial segments occupied by at least one fish $|m_{x,t}|$ by the total number of fish in the trial.

To determine how the escape wave travelled through the group, we identified the maximum distance $w(t) = \max_x(\phi_{x,t} < 0)$, where all the fish in a segment faced away from the stimulus. We call this point $w(t)$ the 'wavefront'. From this, we could calculate the speed at which progressive regions of the group turned and swam in the opposite direction of the stimulus. Specifically, the speed of the wave at time $t$ is $w(t) - w(t-1)$. The escape wave typically occurred in the first half of the arena (up to 845 mm, arc distance, from the stimulus), and therefore the quantification of its speed was limited to segments in this area (figure 1*a* and electronic supplementary material, figures S2*a*–*d*). Only the $n = 28$ trials in which we could determine the location of the wavefront in at least five of these segments were kept for analysis. Once we had identified the time at which the wavefront passed through each segment, we then determined how far the wavefront had travelled from the stimulus on each frame and calculated how many segments it had travelled since the last frame.

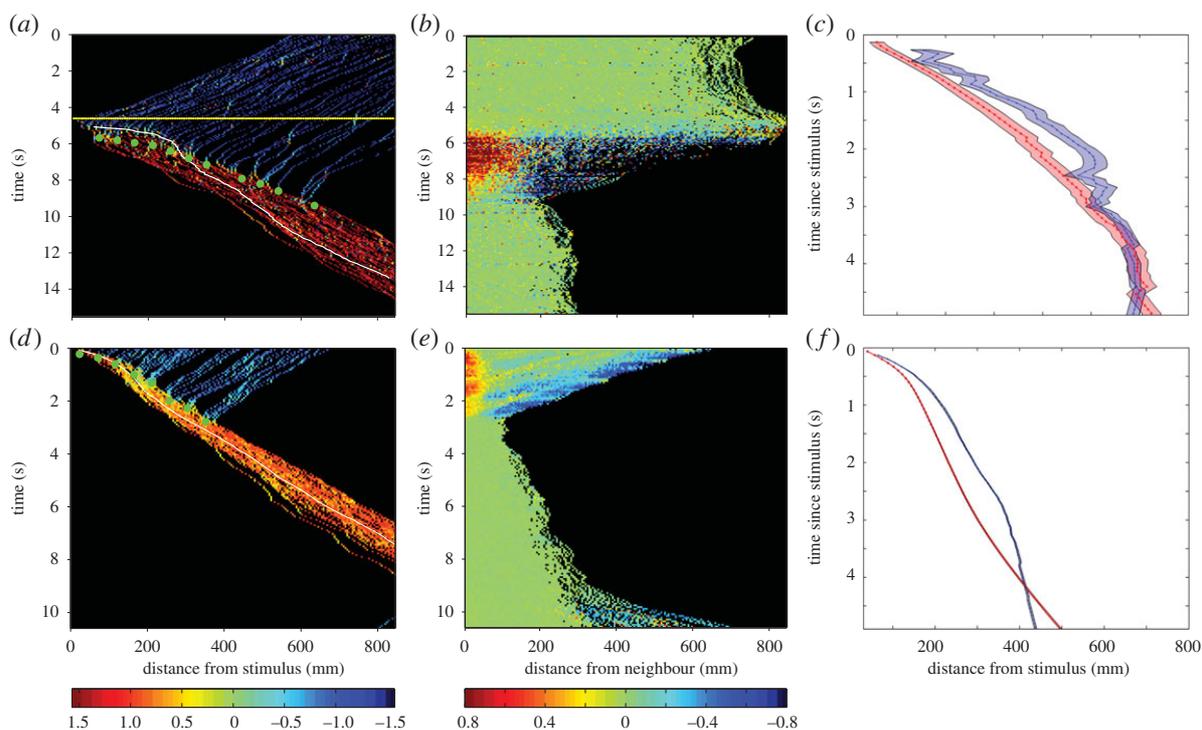

**Figure 1.** Examples of the dynamics of information transfer in an experimental trial with 51 fish (*a*–*c*) and a model simulation (*d*–*f*) with 51 particles. (*a*) Each fish's angular coordinate (in a polar coordinate system with the centre of the arena as a pole and the radius going through the stimulus position as a polar axis) is represented by a colour-coded point. The colour bar below this figure and in (*d*) indicates values in radians, representing a fish's angle relative to the radius $\chi$: individuals in deep blue are moving perpendicularly to the radius and moving in the CW direction, fish in red are oriented perpendicular to the radius but moving anti-CW, and fish in green are moving in a parallel direction to the radius (see Material and methods for details). As the trial progresses in time (*y*-axis), the fish get closer to the stimulus until the stimulus enters the arena (yellow horizontal line). A clear escape response develops, with all individuals moving away from the stimulus. The green circles indicate the position of the escape wave, defined as 10-degree sectors shifting for the first time from a negative to a positive average alignment. The white line shows the trajectory of the first fish to respond to the stimulus. (*b*) Velocity fluctuations (see Material and methods, see colour bar below this figure and (*e*)) measured as a function of distance during the trial. These are uncorrelated before attack, positively correlated at shorter distances and anticorrelated at greater distances during the attack, but uncorrelated following the attack. The points at which the velocity fluctuations reach zero indicate the group's correlation length. (*c*) Average position of the first responding individual (red) and the average position of the escape wave (blue) averaged across all experiments. (*d*) Colour coding and measurements as in (*a*) but for a model simulation. The model shows the same qualitative dynamics as the experimental example in (*a*). (*e*) Velocity fluctuations in the model simulation, again showing consistent dynamics to the experimental results. Colour coding as in (*d*). (*f*) Average position of the informed individuals and the escape wave as in (*c*), but here for model simulations.

Collective evasion responses can be measured by spatial velocity fluctuations [9,10]. To find these, we first calculated the fluctuations in alignment relative to the mean alignment, i.e.

$$\hat{\phi}_{x,t} = \langle \phi_{x,t} - \phi_{x,tx} \rangle,$$

where $\phi_{x,tx}$ is averaged over all space at time, $t$. We then calculated the correlation

$$C_{d,t} = \frac{\sum_{x,x'} \hat{\phi}_{x',t} \hat{\phi}_{x,t} I(x - x', d)}{\sum_{x,x'} I(x - x', d)},$$

where $I(x - x', d)$ is an indicator function, which is uniform if $x - x'$ is between $d$ and $d - (2\pi/36)$ and zero otherwise. Values of $\hat{\phi}_{x,t}$ which are undefined owing to the absence of fish are omitted from the calculation. Note that, unlike in [9], this calculation is not based on individuals, but on local segments of alignment.

All units are given in millimetres, measured as the distance along the arc length of the middle of the annulus from the stimulus.

## 3.4. Model description

We modelled the dynamics of information transfer we observed in the experiments. The model is based on the Vicsek model [36,37], but modified to take into account the acceleration and deceleration behaviour of fish. We model the movement of fish in one dimension, with position determined by the angular distance, $x_i(t)$, of the fish from the stimulus and speed denoted $v_i(t)$. The model does not attempt to account in details for smooth velocity changes in turning angles, but rather direction changes towards and away from the stimulus. As such, speed and velocity could be used interchangeably in what follows. Distance is measured in units of cm along the centre of the ring within the annulus. Position is determined by

$$x_i(t+1) = x_i(t) + v_i(t+1).$$

To calculate speed $v_i(t+1)$, we first find the neighbours ($j \neq i$):

$$N_i(t) = \left\{ \left| x_i(t) - x_j(t) \right| < r \right\},$$

within a distance $r$ of the focal individual. We then calculate the average speed of the neighbours weighted as follows:

$$\hat{v}_i(t) = (1 - \omega_i)v_i(t) + \omega_i \frac{1}{|N_i(t)|} \sum_{j \in N_i(t)} v_j(t).$$

The parameter $\omega_i$ can be thought of as a weighting of social versus private information. $\omega_i = 0$ implies no response to neighbours in determining speed. In the case that there are no neighbours, i.e. $|N_i(t)| = 0$, then $\hat{v}_i(t) = v_i(t)$.

For all simulated fish, the new speed of the fish is then determined by

$$v_i(t+1) = (1-d)\hat{v}_i(t) + d\,\text{sign}(\hat{v}_i(t)) + \epsilon_i(t).$$

The parameter $d$ determines the rate at which speed is dampened to reach the cruise speed after a perturbation, whereas the noise term $\epsilon_i(t)$ models both the variation in the fish movements and provides a density-dependent repulsion. Specifically:

$$\epsilon_i(t) = \frac{\eta_1}{2} u_i^{(1)}(t) + \frac{\eta_2}{2} u_i^{(2)}(t) \frac{|N_i(t)|^3}{T^3 + |N_i(t)|^3},$$

where $u_i^{(1)}(t)$ and $u_i^{(2)}(t)$ are uniformly distributed random variables in a range $[-1, 1]$. The parameters $\eta_1$ and $\eta_2$ control the magnitude of the noise and $T$ determines the density at which collective motion starts to break down. This stochastic repulsion ensures that groups do not become too densely packed. Most of the presented results are based on $\eta_1 = 1$ and $\eta_2 = 0$, i.e. density-independent noise. In all cases, we repeated the analyses with $\eta_1 = 1.4$ and set $\eta_2 = 1.25$ finding no qualitative difference in the results.

The speed-related parameters can be measured directly from the data. For undisturbed and uninformed fish in a group, $s = 0.124 \text{ m s}^{-1}$ is the mean speed for fish swimming in a group (electronic supplementary material, figure S3). The damping parameter $d$ can be fitted from the isolated fish escape profile (electronic supplementary material, figure S3). To perform the fitting, we first identified the reaction time at which speed was maximized, to be one-third of a second. We then performed linear regression of

$$\log(s_{\min} v(t) - s_{\min}) = \log(s_{\text{fright}} - s_{\min}) - dt,$$

where $t = 0$ at the point of maximum speed and $s_{\text{fright}}$ and $s_{\min}$ is the (fitted) level of maximum and minimum speed at that point. This fitting gave $d = 0.0819$.

To investigate whether this model could reproduce the escape waves, we first looked at the response of $N = 51$ individuals (to match experimental group sizes with initial positions chosen uniformly at random between 0 and 310 cm. A proportion $p = 0.1$ of individuals nearest to the stimulus is informed of its existence and initially move rapidly away from it. We call these individuals 'informed'. The other $(1 − p)N = 46$ individuals, which are further away from the stimulus, initially move towards it and are named 'uninformed'. We liken this to other decision-making models that similarly investigate the roles of informed and uniformed individuals during group decision-making [38,39]. For informed individuals, we set $v_i(0) = −2.19 v_{max}$, whereas for uniformed individuals, we set $v_i(0) = v_{max}$, so that they move in the opposite direction.

The other model parameters are the social weighting $\omega_i$, the magnitude of the noise $\eta$ and the range of interaction $r$. In the simulations, we assumed that $\omega_i = 0.4$ for uninformed fish. We further assume that $\omega_i$ is initially 1 for informed/frightened fish, but for frightened fish $\omega_i$ decreases with the same rate constant $d$ as used for the speed. Specifically, for informed fish, we use $\omega_i(0) = 1$ and $\omega_i(t + 1) = (1 − d)\omega_i(t) + 0.4d$ for $t > 0$. We set $r = 3.83$ cm, which is approximately two fish body lengths.

We used the model to investigate how the ability of individuals within a group to respond to a threat depends on the group's density. We set up the simulation with groups of between $N = 5$ and $N = 150$ particles, and with densities varying between 0.03 and 15 particles per cm around the ring's circumference. The initial width of the group, defined as the arc length from the group's leading edge to the group's trailing edge, ranged between 5 and 330 cm. The 10% of informed particles at the front of the group were initially given a higher speed (0.27 m s$^{-1}$, ≈2.19 times the maximum speed of the remaining uninformed group members). Uninformed individuals, who constituted the remainder of the group, were initially given a cruise speed of 0.124 m s$^{-1}$ and travelled in an opposite direction to their informed counterparts.

## 4. Results

### 4.1. Escape waves

Following the simulated attack, we observed a rapid wave of individuals turning away from the stimulus in the majority of trials (figure 1a and electronic supplementary material, figure S2a–d) with groups becoming denser (electronic supplementary material, figure S1b). Across the 39 trials, an average of $84 \pm 16\%$ (±s.d.) of group members changed direction in response to the stimulus. There were five trials in which more than 50% of group members failed to change direction. There was no clear relationship between these five failures and either the group size or the density of the group immediately before the attack (figure 3c).

If a group turns in unison, fluctuations in velocity of individuals should be positively correlated for nearby neighbours, but negatively correlated for distant neighbours. Before the stimulus entered the arena velocity fluctuations were uncorrelated, even at very short distances (figure 1b). This implies that the fluctuations were unaffected by the shape of the arena or any other external factor. Following the attack, the velocity correlations between near neighbours are positive, but decrease monotonically with distance. The point at which the correlations pass from positive to negative is known as the correlation length. Directly after the stimulus the correlation length is proportional to the width of the group moving away from the stimulus. The correlation increases to a maximum when half the fish have turned and then decreases to zero as the remainder of the fish also change direction. These observations were similar across trials (electronic supplementary material, figures S2e–h) and are consistent with a single escape wave travelling through the group.

To quantify how fast the escape wave propagated, we measured how the wavefront changed position through time. The wavefront is defined as the point at which individuals within a region of the group switched from a negative to a positive average alignment after the stimulus was released (green dots in figure 1a and blue line in figure 1c). Following the first second of the stimulus entering the arena, escape waves had a relatively constant speed, although appear to slow down as the wave progresses through the entire group (figure 1c). The initially faster speed of the escape wave is probably owing to individuals responding directly to the threat, and cannot be firmly attributed to an increase in speed of the escape wave at initial perturbation. Group size was not correlated with either an increase or decrease in the average wave speed over the course of the whole trial ($r = 0.072$, $n = 28$, $p = 0.71$) or within the immediate aftermath (2 s) of the stimulus being released ($r = −0.0003$, $n = 25$, $p = 0.99$). The individual that was first to respond to the threat tended to be closest to the stimulus. This individual rapidly accelerated and travelled into the group, later slowing to a speed similar or lower than that of the escape wave (figure 1a,c).

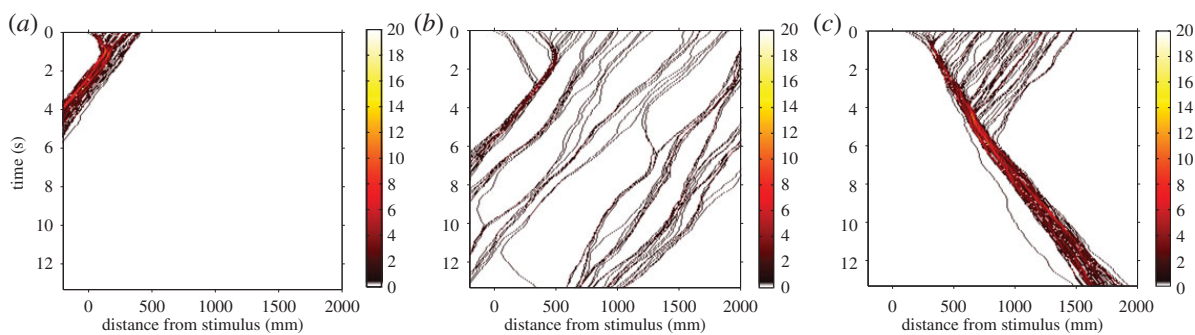

**Figure 2.** How density affects the collective evasion response. Example simulations for 60 individuals spread over (*a*) 42 cm, (*b*) 294 cm, and (*c*) 147 cm. In the simulations, the six individuals nearest to the stimulus were informed and initially moved away from the stimulus. The remaining 54 were uninformed and moved towards the stimulus at normal speed. The 'temperature' of the heat map shows the density of particles (individuals cm$^{-1}$).

### 4.2. Speed changes in groups and isolated fish

The proportion of fast-moving individuals changed immediately after the stimulus was presented: $8.14 \pm 11.0\%$ (mean ± s.d.; electronic supplementary material, figure S4) of individuals moved at more than two times the average initial speed of fish in the first second following the attack, whereas in the second before the attack, less than 1% of fish reached these speeds (electronic supplementary material, figure S4). Similar speed changes were seen in experiments where we startled single fish under the same conditions as the group. While isolated fish had slower speeds than those in groups, they also showed a characteristic increase in swim speed in response to an attack, followed by a reduction in speed to levels close to those before the attack (electronic supplementary material, figure S3).

### 4.3. Modelling the escape wave

We used the model to test the hypothesis that speed changes combined with local interactions facilitate the escape wave. In the model 10% of individuals nearest the front of the group were assumed to be 'informed' and moved more rapidly away from the stimulus, whereas 90% were uninformed, moving more slowly towards the stimulus. Our model was sufficient to reproduce the wave of turning (figure 1*d*) and the spatial fluctuations (figure 1*e*), as well as the speed of the individuals that spotted the stimulus and the resulting escape wave (figure 1*f*).

Density affected whether the group could successfully respond to the threat. At very high densities, when the width of the group was less than 22 cm, nearly all individuals interacted with each other and the group took the direction of the uninformed majority, failing to change direction (figure 2*a*). For densities lower than about 0.3 particles per cm, random fluctuations in individuals' movement caused spontaneous changes in direction and a failure to turn away from the threat (figure 2*b*). At intermediate densities, the wave travelled through the entire group, even when the group was very large (figure 2*c*). There was a wide range of intermediate densities where the entire group turned away from the threat (figure 3*a*).

An increase in speed by informed individuals was essential for allowing the group to reliably change direction. We compared simulation results where the informed individuals increased their speed as we found in the experiments (i.e. the standard value of $v_i(0) = -2.19 v_{max}$) with simulations in which the informed individuals moved at the same speed as the uninformed individuals (i.e. $v_i(0) = -v_{max}$). In this latter case, the group changed direction in at most 50% of simulation runs, compared with over 80% when informed individuals increase their speed (electronic supplementary material, figure S5).

### 4.4. Density and individual escape probability

As each individual in the group should attempt to minimize its chances of being the selected prey item, the group's probability of turning is not an adequate measure of an individual's survival chances. A more appropriate measure is an individual's distance from a potential threat relative to that of its group members [24,25]. In the simulations, we investigated whether the individual nearest the stimulus, 1 s after it was presented, was one of the informed individuals (initially at the front of the group and detecting the predator) or one of the uninformed individuals (initially further back but not detecting the predator). The probability that an informed individual was nearest the threat varied with group density, with uninformed individuals at greatest risk in densely packed groups and informed individuals at most risk in sparse groups (figure 3*b*).

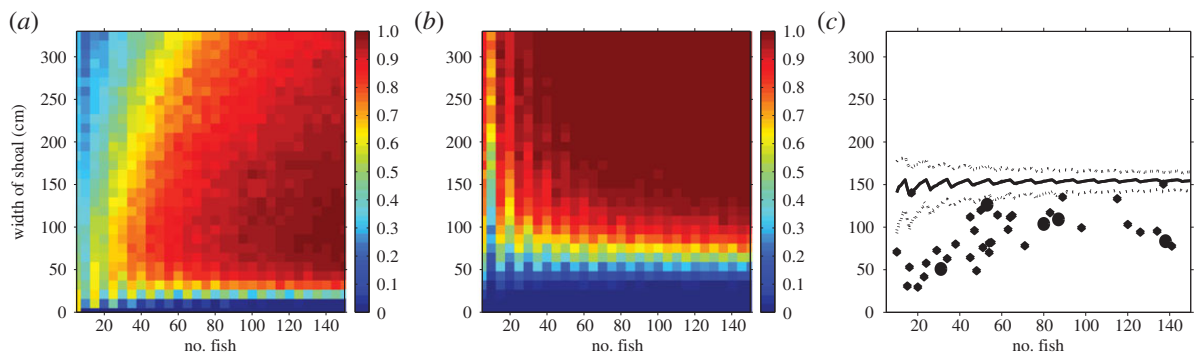

**Figure 3.** The role of density in determining whether a group changes direction in response to a stimulus and which particles escape furthest from the stimulus. (*a*) Proportion of simulations in which all individuals changed direction. (*b*) Proportion of simulations in which the individual nearest to the stimulus 1 s after it appeared was an informed individual. (*c*) Actual width of fish shoals in the experiment before the stimulus was presented (diamonds and circles). These are typically below the expected width under the assumption that the fish are randomly distributed in the ring (average given by solid line, 5 and 95 percentiles given by dotted lines). The black circles represent the five groups that failed to turn.

The typical widths adopted by the real fish groups just prior to the stimulus were significantly lower than if the individuals were distributed uniformly at random around the ring, indicating that individuals in the shoal regulated their density (figure 3*c*). Moreover, for shoals of 50 or more fish the width of the group was around 100 cm, regardless of number of individuals. This value can be compared with 90 cm which, in the model, was the width at which informed individuals were equally likely as uninformed individuals to be closest to the predator 1 s after the attack (figure 3*b*). This observed width was also close to the value that maximized the probability of the group switching direction (figure 3*a*).

### 4.5. Escape waves in larger groups

Naturally occurring shoals of Pacific blue-eyes can reach hundreds of individuals [40]. Observations of other species that form groups consisting of thousands individuals show escape waves passing through the entire group [2,10]. To test whether the mechanisms captured by our model are sufficient to account for escape waves in larger shoals, we simulated 4000 individuals in a group of width 7 m and in which only 0.5% of the individuals were informed (figure 4*a*). In 512 such simulations, the shoal always changed direction and moved away from the stimulus. Further simulations reveal that when density is fixed at a level that facilitates wave propagation, the probability that a group changes direction is independent of the number of individuals (figure 4*b*). The change of direction occurs, because the initial change in speed by the informed individuals results in a local increase in density. As this denser group meets individuals travelling in the opposite direction, it progressively changes their direction and continues to become denser, leading to a snowball effect and a direction change for the entire shoal.

Global interaction rules could not achieve efficient information transfer, because the direction of the 90% of uninformed individuals dominates the 10%. Neither does the result depend on whether interactions are topological and metric (electronic supplementary material, figure S6). The key ingredient to these escape waves, therefore, is an initial increase in individuals' speed leading to a local density increase that propagates through the entire group.

## 5. Discussion

Escape waves are a ubiquitous feature of swarming animal groups under attack, but until now, the mechanisms behind the initiation and spread of these waves remained unknown. We have identified one such mechanism in a laboratory experiment. In our study, the key ingredient to escape wave initiation is a 180° change in direction and an increase in speed by a small number of informed individuals. The increase in density results in a wave of turning that allows information to spread across the entire group.

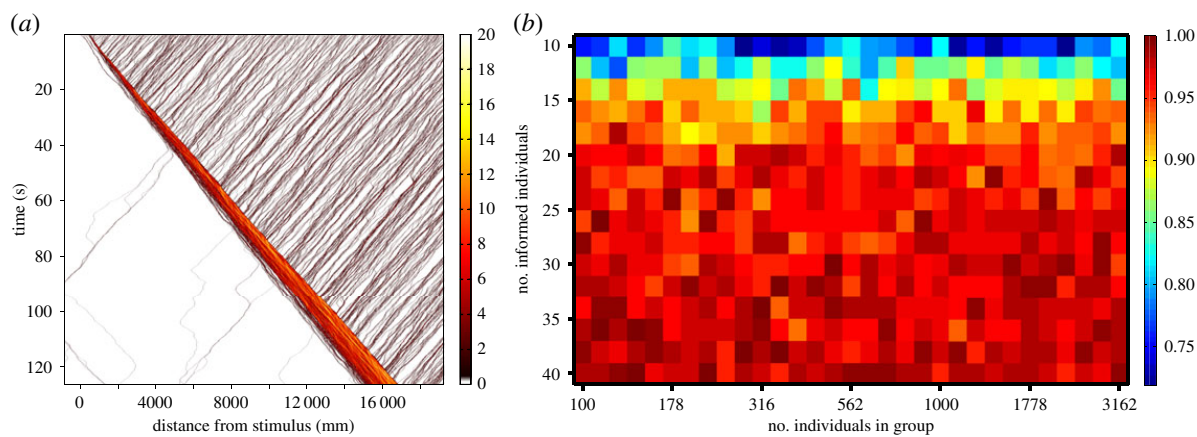

**Figure 4.** Density waves pass through arbitrarily large groups. (a) An example simulation with 4000 individuals spread over 7 m with 20 informed individuals. The 'temperature' of the heat map shows the density of particles (individuals cm$^{-1}$). (b) Escape probability is independent of number of fish when density is held constant at 1.30 individuals cm$^{-1}$. The 'temperature' of the heat map is the proportion of 128 simulation runs for each parameter set in which the entire group turns.

The initiation of the wave is due to individuals changing direction in response to the speed changes of neighbours. There are two reasons why individuals should be highly sensitive to the speeds of other individuals. First, processing information about speed requires limited perception and low cognitive demands. Individuals using a threshold-mediated response to speed information can quickly detect a fast-moving individual among a background of slower-moving group members [41]. Such responses are widespread across taxa [41], and speed changes are commonly observed in bird, insect, crustacean and fish collectives under attack [1–3,12,21,22,42]. This leads to the second reason for using speed information; fast-moving individuals are likely to hold valuable information on the locations of food or predators. As discussed by Lemasson et al. [41], responding to the differential motion of other individuals may also reduce the chances of type 1 errors (false information cascades), because individuals are unlikely to substantially increase their speed in the absence of a threat, and further reduce the chances of type 2 errors (failed responses to real threats). Hence, using the speed changes of individuals appears to be a robust mechanism in allowing accurate information to spread with low cognitive demands to individuals in these groups.

Our model shows that a speed-based mechanism can initiate an escape wave that propagates across the entire group, even after the initial speed changes have disappeared. Waves can spread across arbitrarily large group sizes, because fast-moving individuals initially create a denser band of individuals that influences the direction of others as it moves towards them. Unlike, for example, 'streaker bees' that fly through a honeybee swarm at higher speeds and lead the group [43,44], the initially 'informed' individuals quickly return to their usual swimming speed. The wave continues, because the group has become denser as this wavefront moves away from the source of attack. In contrast to leadership models [38,39], the wave continues to move through the group even when the initially informed individuals have 'forgotten' that they initiated the direction change. This leadership-free form of collective response could prove extremely effective in the case of multiple attacks on a school, because only the most recent attack is 'remembered' by the group.

A speed-initiated density wave could prove a general mechanism for escape wave patterns in other species. In our experiments, as the group turns away from the stimulus, the velocity correlation length increases to become half the length of the entire group, before decreasing again to zero. This could potentially explain observations in starlings and other fish species that velocity correlations scale with group size [9,10]. If density waves pass backwards and forwards within a flock or school, either in response to a predator or owing to intrinsic fluctuations, then these would produce the type of correlation patterns observed in field experiments.

We should, however, be cautious in attributing all animal escape waves to the mechanism we have identified here. The mean speed of the escape wave found in our study (0.29 m s$^{-1}$) is slower than those previously measured in other fishes ([34]; 4.1–10.3 m s$^{-1}$, [12]; 11.8–15.1 m s$^{-1}$). The fish we used were typically smaller than in these other studies (2–3 cm in our study; 18–23 cm in [34]: and 5.3 ± 0.5 cm [12] (±1 s.d.) measured from 20 fish in fig. 16 of [12, p. 84]), and this will clearly affect the speed of the escape wave simply through differences in individual's maximum potential speeds. We also note that the strength of the stimulus could affect the speed of escape responses and the distance it travels. A reliance on visual information, for example, has been suggested to slow down the speed of the escape wave [34]. But despite these differences, it remains clear that the mechanism behind escape wave propagation in our study is different to that in Radakov's work [12]. In both our experiments and model, the escape wave travelled at approximately the same speed as individual fish, with the densest part of the wave being slightly ahead of the trailing edge of the group. These dense waves are often observed in natural animal groups [2,10]. In Radakov's study [12], no dense band of individuals is observed, and instead, the wave consists of individuals turning away from the threat at progressively greater distances from the stimulus. Other mechanisms, possibly similar to those involved in the Mexican wave [45], may facilitate the propagation of escape waves in different species.

Theoretical work on self-propelled particle models [5,46], coupled with a statistical mechanics approach to analysing moving animal groups [11,47], may allow us to unify the different types of waves observed across taxa [48]. In our experiment, the fish are effectively confined to one dimension. Such an arrangement provides only one possible type of escape wave; one that travels directly away from the stimulus. Tu *et al.* [49] use a model to show that moving particles in higher dimensions produce waves in which direction is propagated smoothly. Their model, which also produces 'sound' waves in one dimension, may be the link between the 'fluid' turning waves observed in starlings [11] and our experiments, with dimension providing the determining parameter.

Collective responses performed by unrelated individuals pose an evolutionary conundrum [50]. Why should individuals organize into a state that maximizes information transmission through the entire group, when each individual's survival depends on its own access to information? In our model, we found that the interests of the group and the individual can be aligned, providing an explanation for the evolution of collective evasion responses and information transmission. Both an individual's escape probability and group information transmission are maximized at densities similar to those that the fish adopt in our experiment. As informed individuals increase their speed in response to the attack, they penetrate the group's edge, mixing with their uninformed group members, and presumably diluting their individual risk of capture. The classical selfish herd framework states that individuals should adjust their positions within groups owing to differential predation risk [25]. Our work suggests that individuals can adjust their position dependent on other factors such as hunger or energetic needs [51] without suffering higher predation risk. Alternatively, individuals may be unaware of their relative position in the group and therefore be unable to induce position-dependent escape strategies that could maximize their survival chances. Clearly, it will now be important to investigate these rules under different conditions such as the distance from the threat and the timescale of the attack, like others have done [52].

Ethics statement. All experiments were conducted in accordance with Sydney University's Animal Ethics Committee (ref. number: L04/6-2009/3/5083).

Data accessibility. Data accompanying this paper can be found at datadryad.org (doi:10.5061/dryad.3f1tm).

Acknowledgements. We thank N.O. Handegard for providing the tracking software and Björn Rogell for statistical advice and two anonymous referees and the editor for their useful suggestions.

Funding statement. The work was funded by a start-up grant from the University of Sydney awarded to A.J.W.W., an ARC Discovery Project awarded to A.J.W.W. and J.B., an ARC Future Fellowship awarded to J.B. and an ERC grant awarded to D.J.T.S.

Author contributions. J.H.R., A.J.W.W. and F.H. designed the study. J.H.R. performed the experiments. J.H.R., J.B., F.H. and D.J.T.S. analysed the data. J.H.R., J.B., A.J.W.W. and D.J.T.S. wrote the paper. All authors gave final approval for publication.

Competing interests. We have no financial or non-financial competing interests.